\documentclass[amsmath,amssymb,aps,pra,twocolumn]{revtex4-1}
\usepackage{bm}
\usepackage{amsfonts}
\usepackage{amssymb}
\usepackage{graphicx}

\begin{document}

\title{Comment on "Uncertainty relations in terms of the Tsallis entropy"}
\author{Iwo Bialynicki-Birula}\email{birula@cft.edu.pl}
\author{{\L}ukasz Rudnicki}\email{rudnicki@cft.edu.pl}
\affiliation{Center for Theoretical Physics, Polish Academy of Sciences\\
Al. Lotnik\'ow 32/46, 02-668 Warsaw, Poland}

\begin{abstract}
We prove that the inequality used recently by Wilk and W{\l}odarczyk [Phys. Rev. A {\bf 79}, 062108 (2009)] to find a better lower bound in the uncertainty relations for the R\'enyi entropies is invalid. Thus, the problem of improving the bound given in our paper [Phys. Rev. A {\bf 74}, 052101 (2006)] remains unsolved.
\end{abstract}
\maketitle

It has been shown in our earlier paper \cite{ibb} that the uncertainty relation restricting the values of the R\'enyi entropies for position $H^{(x)}_{\beta}$ and momentum $H^{(p)}_{\alpha}$ has the following form:
\begin{eqnarray}\label{reur}
H^{(p)}_\alpha+H^{(x)}_\beta\geq-\frac{1}{2}\left(\!\frac{\ln \alpha}{1-{\alpha}}+\frac{\ln \beta}{1-\beta}\!\right)-\ln\left(\!\frac{\delta x\delta p}{\pi\hbar}\!\right)\!,
\end{eqnarray}
where $\delta x$ and $\delta p$ determine the bin sizes for position and momentum. The parameters $\alpha$ and $\beta$ are assumed to be positive and they are constrained by the relation:
\begin{eqnarray}\label{relation}
\frac{1}{\alpha}+\frac{1}{\beta}=2.
\end{eqnarray}
The R\'enyi entropy is defined in a standard way,
\begin{eqnarray}\label{renyi}
H_\alpha=\frac{1}{1-\alpha}\ln\left(\sum p_k^\alpha\right).
\end{eqnarray}
In the limit, when $\alpha\to 1$ and ${\beta}\to 1$, this uncertainty relation reduces to the uncertainty relation for the Shannon entropies \cite{ibb1}
\begin{eqnarray}\label{seur}
H^{(p)} + H^{(x)}\geq -\ln\left(\frac{2\delta x\delta p}{e h}\right).
\end{eqnarray}
The bounds in Eqs.~(\ref{reur}) and (\ref{seur}) are certainly not saturated. Even worse, they become negative for large values of $\delta x\delta p/\hbar$, whereas the left hand side is always positive. Therefore, an improvement of these bounds is highly desired.

The bound proposed by Wilk and W{\l}odarczyk in the uncertainty relation (\ref{seur}) has the form
\begin{eqnarray}\label{nb}
H^{(p)} + H^{(x)}\geq-\ln\left(\frac{2}{e}\frac{\delta x\delta p}{h+\delta x\delta p}\right).
\end{eqnarray}
This lower bound has the attractive property of being always positive but it cannot be correct because in the limit $\delta x\delta p\to \infty$ it does not approach 0. As a matter of fact, if one follows the procedure proposed in \cite{ww}, the lower bound in (\ref{nb}) is determined only up to an additive constant. This constant depends on an arbitrary choice of the length $2a$ of the line segment used in Eq.~(25) of Ref.~\cite{ww} and is given by the following formula:
\begin{align}
-\ln\frac{2 a^2}{e}=1-\ln 2-2\ln a.
\end{align}
The authors have chosen $a=1$, but any other choice is equally good. We would like to stress that in any case the bound should always tend to 0, when the sizes of the bins $\delta x$ and $\delta p$ tend to $\infty$. We shall illustrate this fact with a simple example of a Gaussian wave function:
\begin{align}\label{gx}
\psi(x)=\left(\frac{1}{\pi\sigma^2}\right)^{1/4}\!\!\!\!e^{ip_0(x-x_0/2)/\hbar}
\exp\left[-\frac{(x-x_0)^2}{2\sigma^2}\right],
\end{align}
and its Fourier transform
\begin{align}\label{gp}
{\tilde\psi}(p)=\left(\frac{\sigma^2}{\pi\hbar^2}\right)^{1/4}\!\!\!\!e^{-ix_0(p-p_0/2)/\hbar}
\exp\left[-\frac{\sigma^2(p-p_0)^2}{2\hbar^2}\right].
\end{align}
Let us choose now two infinitely large bins $(-\infty,0)$ and $(0,\infty)$ for both position and momentum. The left hand side of the uncertainty relation (\ref{nb}) for this choice of the wave function is equal to
\begin{align}\label{erf}
H^{(p)} + H^{(x)}&=2\ln 2-(1-{\rm Erf}(\delta))\ln(1-{\rm Erf}(\delta))\nonumber\\
&-(1+{\rm Erf}(\delta))\ln(1+{\rm Erf}(\delta)),
\end{align}
where $\delta=\sqrt{x_0p_0/\hbar}$ and we have chosen, for simplicity, $\sigma=x_0/\delta$. For large values of $\delta$, the expression on the right hand side of (\ref{erf}) tends to 0. The limit $\delta\to\infty$ means that we localize the Gaussians (\ref{gx}) and (\ref{gp}) so far to the right that its presence in the left bin is negligible.

The existence of a counterexample means that there is an error in the proof. We shall now explain where the authors made a mistake. In the original derivation given in \cite{ibb} a crucial role is played by the inequality (Eq.~(21) of Ref.~\cite{ibb}) which was rewritten in \cite{ww} in the form
\begin{eqnarray}\label{bb}
-\left(\sum_k p_k^\alpha\right)^{1/\alpha}\geq-\eta(\alpha,\beta)\left(\sum_l x_l^\beta\right)^{1/\beta},
\end{eqnarray}
where
\begin{eqnarray}\label{eta}
\eta(\alpha,\beta)=\left(\frac{\beta}{\alpha}\right)^{1/2\alpha}\left(\frac{2\beta\delta x\delta p}{h}\right)^{1-1/\alpha},
\end{eqnarray}
and the numbers $p_k$ and $x_l$ are the probabilities to find the momentum and the position in the $k$-th and $l$-th bin, respectively. As was shown in \cite{ibb}, this inequality follows from the Babenko-Beckner inequality for the $p$ and $q$ norms of a function and its Fourier transform. It is important to stress that the inequality (\ref{bb}) holds only because the probabilities $p_k$ and $x_l$ are defined in terms of wave functions that are {\em connected by Fourier transformation}.

The decisive role in the derivation presented in \cite{ww} is played by the following inequality (Eq.~(31) of Ref.~\cite{ww}) which is patterned after our inequality (\ref{bb}):
\begin{eqnarray}\label{ww}
-\left(\sum_k p_k'^\alpha\right)^{1/\alpha}\geq-\eta(\alpha,\beta)\left(\sum_l x_l'^\beta\right)^{1/\beta},
\end{eqnarray}
where
\begin{eqnarray}\label{neweta}
\eta(\alpha,\beta)=\left(\frac{\beta}{\alpha}\right)^{1/2\alpha}(2\beta\delta t_x\delta t_p)^{1-1/\alpha}.
\end{eqnarray}
However, the authors seem to have missed the crucial point in our derivation. They failed to notice that the probabilities $p_k'$ and $x_l'$ that enter their inequality are {\em not related} by the Fourier transformation. Of course, one may always change arbitrarily the integration variables in the correct inequality (\ref{bb}). However, one should also {\em change the integration measure} and this has not been done correctly in Ref.~\cite{ww}.

To make this Comment complete, we invoke the full definitions of $p_k'$ and $x_l'$. As the first step, the authors compactify the infinite ranges of $x$ and $p$ by the following change of variables:
\begin{eqnarray}\label{new}
t_x=\frac{x}{|x|+s_x},\quad\quad t_p=\frac{p}{|p|+s_p}.
\end{eqnarray}
The new variables vary between -1 and 1 and the parameters $s_x$ and $s_p$ are free to choose. Next, they define $p_k'$ and $x_l'$ with respect to the bins corresponding to these new variables, i.e.,
\begin{subequations}\label{newprime}
\begin{eqnarray}
p_k'=\int_{k\delta t_p}^{(k+1)\delta t_p}\!\!dt_p\,{\tilde\rho}[p(t_p)]\frac{s_p}{(1-|t_p|)^2},\\
x_l'=\int_{l\delta t_x}^{(l+1)\delta t_x}\!\!dt_x\,{\rho}[x(t_x)]\frac{s_x}{(1-|t_x|)^2},
\end{eqnarray}
\end{subequations}
where ${\tilde\rho}(p)$ and $\rho(x)$ are the original probability densities and $\delta t_p$ and $\delta t_x$ are the bin sizes in the new variables. Extra factors arise from the change of variables. Since after the change of variables the wave functions are no longer related by the Fourier transformation, there is no reason to expect that the inequality (\ref{ww}) holds. To drive the point home, we give a straightforward counterexample proving that the inequality (\ref{ww}) is indeed wrong.

To this end, let us consider an arbitrary normalized even function $\psi(x)$ localized on a small segment of the real axis,
\begin{eqnarray}\label{loc}
|x|\leq a=\frac{s_x\delta t_x}{1-\delta t_x},
\end{eqnarray}
where the parameters $s_x>0$ and $1>\delta t_x>0$ have the same meaning as in Ref.~\cite{ww}. For example, we could use the following function $\psi(x)$
 \begin{eqnarray}\label{const}
\psi(x)=\frac{\theta(a-|x|)}{\sqrt{2a}}.
\end{eqnarray}
Its Fourier transform is
\begin{eqnarray}\label{ft}
{\tilde\psi}(p)=\sqrt{\frac{\hbar}{a \pi}}\frac{\sin(a p/\hbar)}{p}.
\end{eqnarray}
This function occupies two bins of size $\delta t_x$ in the $t_x$ variable. The Fourier transform (\ref{ft}) of this function is also an even function.

Next, we choose in the inequality (\ref{ww}) the value of $\delta t_p=1$. This means that we divided the whole range $(-1,1)$ of $t_p$ also into two segments. In the notation of \cite{ww}, the value of $k_{\rm max}$ is equal to 1. Under these assumptions the probabilities $p_k'$ and $x_l'$ are all equal to 1/2 and the inequality (\ref{ww}) reads
\begin{eqnarray}\label{viol}
-\left(2^{1-\alpha}\right)^{1/\alpha}\geq-\left(\frac{\beta}{\alpha}\right)^{1/2\alpha}
\!\!\!\!\!\!\left(2\beta\delta t_x\right)^{1-1/\alpha}\left(2^{1-\beta}\right)^{1/\beta}\!\!,
\end{eqnarray}
or equivalently
\begin{eqnarray}\label{or}
\delta t_x\geq \frac{1}{8\alpha}(2\alpha-1)^\frac{2\alpha-1}{2\alpha-2}\geq 1/4.
\end{eqnarray}
Choosing a sufficiently small arbitrary parameter $\delta t_x$ we shall clearly violate this inequality. Thus, the inequality (\ref{ww}) does not hold and all conclusions based on this inequality in Ref.~\cite{ww} are not valid.
\acknowledgments This research was partly supported by the grant from the Polish Ministry of Science and Higher Education.

\end{document}